\documentclass[conference,compsoc]{IEEEtran}
\ifCLASSOPTIONcompsoc
  \usepackage{cite}
\else
  \usepackage{cite}
\fi
\ifCLASSINFOpdf
\else
\fi
\usepackage{tikz}
\usepackage{subfigure}
\usepackage{multirow}
\usepackage{amsmath}
\usepackage{lipsum}
\usepackage{threeparttable}
\usepackage{makecell}
\usepackage{titlesec}
\usepackage[ruled,vlined, linesnumbered]{algorithm2e}
\usepackage{algorithmic}
\usepackage{cleveref}
\usepackage{listings}
\usepackage[strict]{changepage}
\usepackage{float}
\usepackage{ulem}
\usepackage{paralist}
\usepackage{threeparttable}
\usepackage{ifthen} 
\usepackage{balance}
\usepackage{tcolorbox}
\usepackage{url}

\newtcolorbox[auto counter]{findingbox}[1][]{
    colframe=titleback,
    colback=white!98!black,
    colbacktitle=titleback,
    title=Finding~\thetcbcounter #1
}

\definecolor{titleback}{RGB}{240,152,140}
\colorlet{punct}{red!60!black}
\definecolor{background}{RGB}{238,238,238}
\definecolor{delim}{RGB}{20,105,176}
\colorlet{numb}{magenta!60!black}
\lstdefinelanguage{json}{
    basicstyle=\footnotesize\ttfamily,
    showstringspaces=false,
    breaklines=true,
    frame=lines,
    captionpos=b,
    backgroundcolor=\color{background},
    literate=
     *{0}{{{\color{numb}0}}}{1}
      {1}{{{\color{numb}1}}}{1}
      {2}{{{\color{numb}2}}}{1}
      {3}{{{\color{numb}3}}}{1}
      {4}{{{\color{numb}4}}}{1}
      {5}{{{\color{numb}5}}}{1}
      {6}{{{\color{numb}6}}}{1}
      {7}{{{\color{numb}7}}}{1}
      {8}{{{\color{numb}8}}}{1}
      {9}{{{\color{numb}9}}}{1}
      {:}{{{\color{punct}{:}}}}{1}
      {,}{{{\color{punct}{,}}}}{1}
      {\{}{{{\color{delim}{\{}}}}{1}
      {\}}{{{\color{delim}{\}}}}}{1}
      {[}{{{\color{delim}{[}}}}{1}
      {]}{{{\color{delim}{]}}}}{1},
}

\newboolean{showrevisions}
\setboolean{showrevisions}{false} 

\ifbool{showrevisions}
    {\newcommand{\newtext}[1]{\textcolor{blue}{#1}}\newcommand{\removedtext}[1]{\textcolor{red}{\sout{#1}}}}
    {\newcommand{\newtext}[1]{#1}\newcommand{\removedtext}[1]{}}
    
\IEEEoverridecommandlockouts

\begin{document}

\title{The Digital Cybersecurity Expert: How Far Have We Come?}
\author{
    \IEEEauthorblockN{
        Dawei Wang\IEEEauthorrefmark{2},
        Geng Zhou\IEEEauthorrefmark{2},
        Xianglong Li\IEEEauthorrefmark{2},
        Yu Bai\IEEEauthorrefmark{2},
        Li Chen$^{*}$\thanks{* Corresponding author.}\IEEEauthorrefmark{2},
        Ting Qin\IEEEauthorrefmark{2},
        Jian Sun\IEEEauthorrefmark{2}, and
        Dan Li\IEEEauthorrefmark{3}
    }
    \IEEEauthorblockA{
        \IEEEauthorrefmark{2}Zhongguancun Laboratory, \IEEEauthorrefmark{3}Tsinghua University\\ 
        \{wangdw, zhougeng, lixl, baiyu, chenli, qingting, sunjian\}@zgclab.edu.cn, tolidan@tsinghua.edu.cn
    }
}

\maketitle

\begin{abstract}

The increasing deployment of large language models (LLMs) in the cybersecurity domain underscores the need for effective model selection and evaluation. However, traditional evaluation methods often overlook specific cybersecurity knowledge gaps that contribute to performance limitations. To address this, we develop CSEBenchmark, a fine-grained cybersecurity evaluation framework based on 345 knowledge points expected of cybersecurity experts. Drawing from cognitive science, these points are categorized into factual, conceptual, and procedural types, enabling the design of 11,050 tailored multiple-choice questions. We evaluate \removedtext{10}\newtext{12} popular LLMs on CSEBenchmark and find that even the best-performing model achieves only 85.42\% overall accuracy, with particular knowledge gaps in the use of specialized tools and uncommon commands. Different LLMs have unique knowledge gaps. Even large models from the same family may perform poorly on knowledge points where smaller models excel. By identifying and addressing specific knowledge gaps in each LLM, we achieve up to an \removedtext{84.18}\newtext{84}\% improvement in correcting previously incorrect predictions across three existing benchmarks for two cybersecurity tasks. Furthermore, our assessment of each LLM's  \removedtext{suitability for}\newtext{knowledge alignment with} specific cybersecurity roles reveals that different models align better with different roles, such as GPT-4o for \removedtext{ISC Security Engineer}\newtext{the Google Senior Intelligence Analyst} and \removedtext{Qwen-2.5-72B}\newtext{Deepseek-V3} for \removedtext{Red Team Security Engineer}\newtext{the Amazon Privacy Engineer}. These findings underscore the importance of aligning LLM selection with the specific knowledge requirements of different cybersecurity roles for optimal performance.

\end{abstract}
\IEEEpeerreviewmaketitle

\section{Introduction}
\label{sec:introduction}

The rapid advancement of large language models (LLMs) has the potential to revolutionize the cybersecurity field, with the concept of a ``digital cybersecurity expert'' gaining traction. As these models become increasingly sophisticated, there is growing interest in their ability to assist or even replace human experts in various cybersecurity tasks. The cybersecurity industry has already begun exploring this possibility, with Microsoft introducing Copilot for Security to proactively detect, investigate, and respond to threats~\cite{copilot_for_security}, and Google launching Gemini in Security to support threat intelligence analysis and streamline security operations~\cite{gemini_in_security}. These developments raise a critical question: \textbf{How far have we come in achieving a digital cybersecurity expert?} Answering this question is crucial for understanding the current capabilities and limitations of LLMs in the cybersecurity domain, which in turn has significant implications for the future of the field. As organizations increasingly rely on these models to support or even replace human experts, it is essential to have a clear understanding of their strengths and weaknesses to ensure the effective and responsible deployment of LLMs in cybersecurity roles.

Recent studies have attempted to evaluate LLMs' capabilities in cybersecurity, which primarily focus on two main areas: their performance on specific security tasks~\cite{alam2024ctibench, srikanth2024evaluating, shafee2024evaluation, liu2024cyberbench, ji2024sevenllm, ullah2024llms, alrashedy2023can, zibaeirad2024vulnllmeval, zhang2024cybench, tian2024debugbench, fang2024large, yang2024evaluation, bhatt2023purple, gong2024well, yang2024seccodeplt, liu2024vuldetectbench} and their understanding of cybersecurity knowledge~\cite{liu2023secqa, tihanyi2024cybermetric, levi2024cyberpal, bhusal2024secure, li2023seceval, alam2024ctibench}. These studies have identified several limitations of LLMs in cybersecurity applications, while offering valuable insights to the community. However, despite these contributions, these works are insufficient to comprehensively assess the knowledge of LLMs in cybersecurity due to the following limitations:

\noindent \textbf{Limitations.}
\begin{inparaenum}[\bf L1-]
    \item \textit{Lack of a comprehensive knowledge framework for cybersecurity experts}: Existing evaluation methods fail to address the fundamental question: what constitutes a cybersecurity expert? These methods often focus narrowly on specific skills or tasks, without establishing a comprehensive framework for the knowledge a cybersecurity expert should possess. As a result, the evaluation questions lack depth and fail to systematically cover necessary areas. Some knowledge domains are overemphasized, while equally important ones are arbitrarily neglected, leading to incomplete and unbalanced assessments.
    
    \item \textit{Inability to identify specific knowledge gaps of LLMs}: Current knowledge-based assessments are coarse-grained, making it difficult to assess LLMs' understanding of specific knowledge points and identify their true knowledge gaps. While some studies~\cite{li2023seceval, tihanyi2024cybermetric} have categorized subdomains within cybersecurity, evaluations within these subdomains lack sufficient detail, limiting their usefulness for model improvement. In task-based assessments, although LLMs' poor performance on certain tasks is apparent, the lack of clear definitions of the required knowledge makes it difficult to identify the causes of failure. This highlights the need for fine-grained evaluation datasets that can provide actionable insights for model enhancement.
    
    \item \textit{Mismatch between question design and knowledge mastery requirements}: Different types of knowledge points require different levels of mastery from cybersecurity experts. For example, knowledge of \textit{HTTP status codes} only requires memorization, while \textit{SSL} requires an understanding of its internal mechanisms, and \textit{Wireshark} requires hands-on proficiency. Each type of knowledge point requires a tailored evaluation approach. However, existing evaluations often use a one-size-fits-all question design, leading to over-emphasis of some areas and insufficient assessment of others, making it difficult to accurately measure LLMs' mastery across different knowledge types.
\end{inparaenum}

To address these limitations, we design a cognitive science-based, fine-grained knowledge assessment framework for cybersecurity experts, called CSEBenchmark. CSEBenchmark uses multiple-choice questions to evaluate LLMs. To accurately depict the knowledge and skills required of cybersecurity experts, we collect three well-known cybersecurity expert roadmaps~\cite{cybersecurity_expert_roadmap,ethical_hacking_roadmap, computer_security_roadmap}, which outline the essential skills and knowledge needed, and consolidate them into a knowledge framework encompassing seven subdomains, including Fundamental IT Skills (FIS), Operating Systems (OS), Networking Knowledge (NK), Web Knowledge (WK), Security Skills and Knowledge (SSK), Cloud Skills and Knowledge (CSK), and Programming Skills and Knowledge (PSK). The entire framework consists of 345 fine-grained knowledge points, providing a comprehensive assessment of LLMs' understanding of these knowledge domains. Given the varying levels of mastery required for different knowledge points, we categorize them based on cognitive science into three types: factual knowledge (to be memorized), conceptual knowledge (requiring understanding of underlying principles), and procedural knowledge (requiring hands-on practice). For each category, we gather targeted materials and design tailored question templates to ensure a comprehensive and accurate evaluation. We use GPT-4-Turbo to generate the questions, followed by 672 man-hours of review and 100 man-hours of corrections, resulting in 11,050 high-quality multiple-choice questions.

We apply CSEBenchmark to \removedtext{10}\newtext{12} popular LLMs, revealing GPT-4o as the overall best-performing model and \removedtext{Qwen-2.5-72B}\newtext{Deepseek-V3} as the top open-source model. However, the overall accuracy of the models is only as high as 85.42\%, indicating room for improvement. We also reveal that LLMs have notable gaps in procedural knowledge, especially in the use of specialized tools and uncommon commands. Additionally, they even struggle with some foundational factual and conceptual points. Notably, different LLMs exhibit unique knowledge gaps, and even larger models from the same family may underperform on certain knowledge points where smaller models excel. By supplementing these knowledge gaps, we successfully enhance their performance across three existing benchmarks~\cite{liu2024vuldetectbench, ullah2024llms, alam2024ctibench} for two cybersecurity tasks, achieving an improvement of up to \removedtext{84.18}\newtext{84}\% in correcting previously incorrect predictions, which validates the reliability of our findings. Finally, we evaluate the job-role \newtext{knowledge} alignment of LLMs based on six real-world cybersecurity roles, demonstrating that LLMs are not yet fully capable of meeting real-world job requirements. Each cybersecurity role reveals unique \removedtext{skill}\newtext{knowledge} gaps within the LLMs, emphasizing the need for role-specific improvements.

\noindent \textbf{Contributions}. Our contributions are summarized as follows:

$\bullet$\textit{New evaluation framework}. We introduce CSEBenchmark, the first cognitive science-based cybersecurity knowledge assessment framework that encompasses 345 fine-grained knowledge points across seven key subdomains critical to cybersecurity experts. This framework offers a comprehensive evaluation of LLMs' understanding of cybersecurity. The benchmark includes 11,050 high-quality multiple-choice questions, with 772 man-hours spent on review and correction, and \$234.5 allocated for question generation. We release our framework~\footnote{https://github.com/NASP-THU/CSEBenchmark} to provide the community with the tools to assess emerging LLMs and conveniently track their progress in mastering cybersecurity expertise.

$\bullet$\textit{New findings}. We evaluate \removedtext{10}\newtext{12} popular LLMs using CSEBenchmark, incurring a total of 1.08 GPU-weeks and costing \$2140.01. The results indicate that current LLMs still fall short of fulfilling the role of a cybersecurity expert, particularly in handling specialized tools and uncommon commands. By addressing these knowledge gaps, we achieve an improvement of up to \removedtext{84.18}\newtext{84}\% in correcting previously incorrect predictions across three existing cybersecurity evaluation datasets, validating the effectiveness of our findings. Lastly, we assess the job-role \newtext{knowledge} alignment of LLMs across six real-world cybersecurity job roles, revealing that LLMs struggle to fully meet these roles' requirements. Different LLMs show varying degrees of suitability, suggesting that model selection should be tailored to specific task demands.

\section{Background and Related Work}
\label{sec:background_related_work}

\begin{figure*}[ht]
    \centering
    \includegraphics[width=0.8\textwidth]{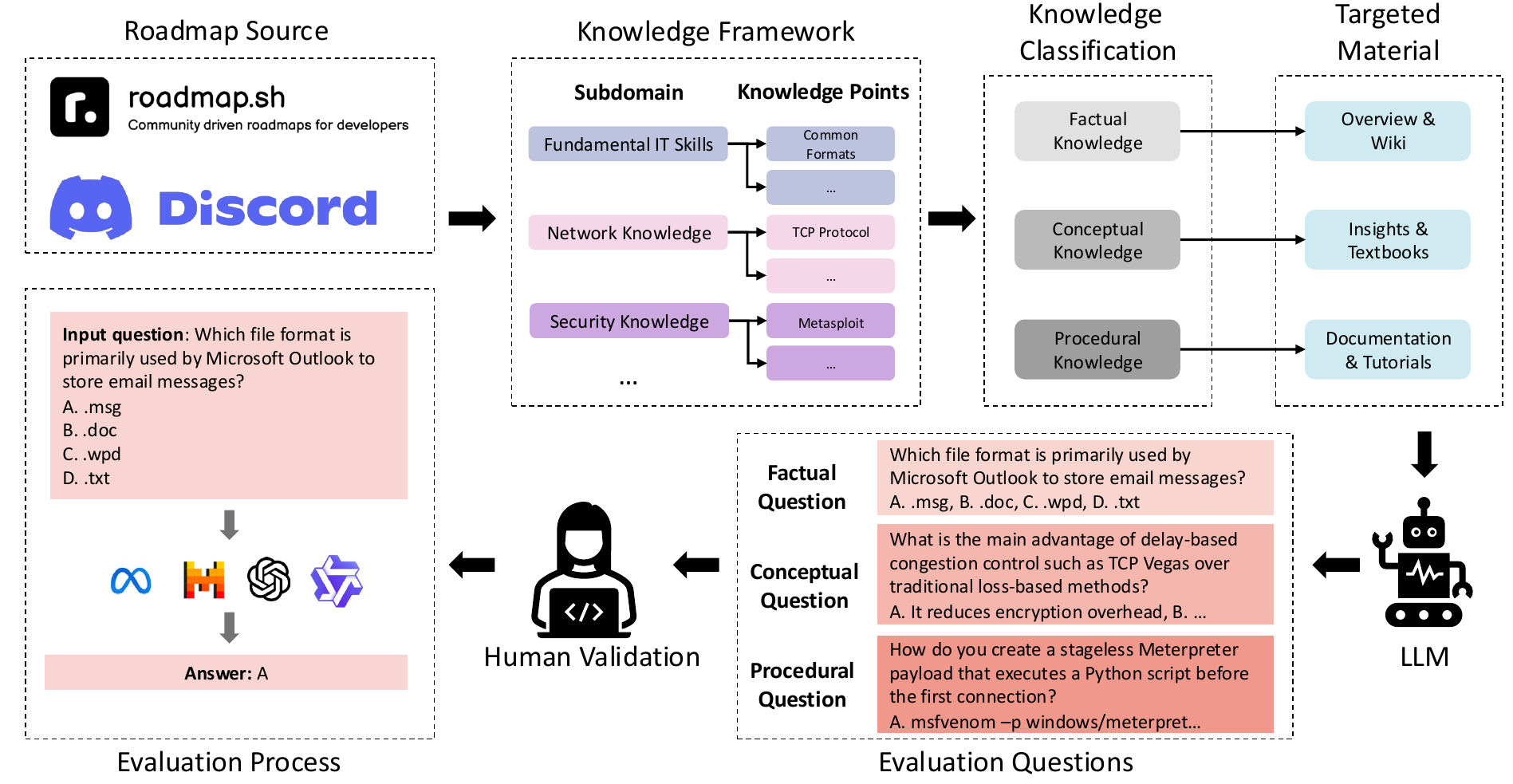}
    \caption{Overview of the construction process of CSEBenchmark.}
    \label{fig:overview}
\end{figure*}

\subsection{Large Language Model}
\label{subsec:background_llm}

Large language models (LLMs) have seen rapid development, leading to significant advancements in natural language processing and understanding. These models, such as OpenAI's GPT series and Meta's Llama, are capable of handling a variety of tasks including translation, summarization, content generation, and question answering. Techniques such as Zero-shot learning enable LLMs to approach new tasks without specific training examples~\cite{wei2021finetuned}, while Few-shot learning allows them to adapt quickly with minimal examples~\cite{brown2020language}. Additionally, Chain-of-Thought (CoT) reasoning enhances complex problem solving by guiding models to break down multi-step tasks logically, yielding clearer and more accurate responses~\cite{wei2022chain}. These capabilities make LLMs highly versatile, finding use in applications such as customer service chatbots, virtual assistants, content recommendation systems, and creative writing. Their flexibility and adaptability have made them useful in business, education, healthcare, and more. Their ability to process and generate human-like text has made them increasingly popular across various fields, sparking interest in their potential to assist or even replace human experts.

In cybersecurity, LLMs have started to demonstrate their value in helping with complex tasks that were traditionally performed by experts. For example, LLMs have been applied to support threat intelligence analysis by gathering, processing, and summarizing threat data from multiple sources, helping analysts identify potential risks more efficiently~\cite{clairoux2024use, kucsvan2024inferring, huangctikg, mitra2024localintel, zhang2024cybench, fieblinger2024actionable}. In incident response, LLMs help by providing real-time recommendations, generating response playbooks, and analyzing incident logs to determine the root cause of security breaches~\cite{hays2024employing, ahmed2023recommending, chen2024automatic, zhang2024automated}. For vulnerability assessment, they help by scanning codebases for known vulnerabilities~\cite{lu2024grace, ghosh2024cve, du2024vul, yang2024dlap}, suggesting patches~\cite{yu2024llm, alrashedy2023can, ahmed2023better, hidvegi2024cigar}, predicting potential weaknesses based on historical data~\cite{deng2023large, wang2024prophet}, and performing reverse engineering to identify hidden or complex vulnerabilities~\cite{hu2024degpt, tan2024llm4decompile, she2024wadec, rong2024disassembling}. Additionally, LLMs are used to automate routine security operations such as reading documentation~\cite{ma2024one, meng2024large, wang2024llmif}, understanding code~\cite{liu2024detecting, nam2024using, zhang2024detecting}, and assisting in vulnerability management~\cite{luo2024cvecenter, liu2024exploring}, which significantly reduces the workload for security teams. Despite these advancements, the question remains: how far have we progressed towards developing LLMs that can fully assume expert roles in cybersecurity?

\subsection{Evaluation of LLMs in Cybersecurity}
\label{subsec:background_evaluation}

Evaluating LLMs involves assessing their capabilities to meet specific standards and effectively perform targeted tasks. These evaluations are generally divided into task-based and knowledge-based assessments. Task-based assessments, on the one hand, evaluate the model's ability to perform cybersecurity-related tasks, such as analyzing threat intelligence, managing vulnerabilities or generating secure code. These assessments typically involve end-to-end tasks framed within real-world security scenarios. For example, in threat intelligence analysis, LLMs are primarily required to analyze real-world threat intelligence reports, assessing their capabilities in named entity recognition, intelligence classification, summarization, and attribution~\cite{alam2024ctibench, srikanth2024evaluating, shafee2024evaluation, liu2024cyberbench, ji2024sevenllm}. Similarly, evaluations in vulnerability management typically provide carefully selected code snippets, requiring LLMs to comprehend code, debug, generate unit tests, identify vulnerabilities, and apply patches to assess their capabilities in each of these areas~\cite{ullah2024llms, alrashedy2023can, zibaeirad2024vulnllmeval, zhang2024cybench, tian2024debugbench, fang2024large, yang2024evaluation, liu2024vuldetectbench}. In secure code generation, LLMs are tasked with generating diverse code, and their capability in secure coding is assessed by evaluating the security of the code they produce~\cite{bhatt2023purple, gong2024well, yang2024seccodeplt}. Although these task-based evaluations intuitively demonstrate model performance across various tasks, they have limitations in identifying the underlying reasons for results due to the lack of quantification of the knowledge needed for each task, making it challenging to conduct a targeted analysis or identify specific weaknesses in the models.

Knowledge-based assessments, on the other hand, gauge a model’s understanding of specialized cybersecurity domains, often through multiple-choice questions (MCQs) generated from relevant materials. For example, SecQA~\cite{liu2023secqa} generates approximately 200 questions from the book \textit{``Computer Systems Security: Planning for Success''} to assess security principles knowledge. CyberMetric~\cite{tihanyi2024cybermetric} and SecEval~\cite{li2023seceval} use 10,000 and 2,126 questions, respectively, drawn from textbooks, documentation, and industry guides to assess expertise across areas such as penetration testing, cryptography, and network security. CTIBench~\cite{alam2024ctibench} generates 2,500 questions from CTI frameworks, regulations, and public resources to evaluate knowledge in cyber threat intelligence. CyberPal.AI~\cite{levi2024cyberpal} builds on CTIBench, SecEval, and other publicly available questions, such as CISSP Assessment Questions and SecMMLU, to evaluate a broader range of LLM knowledge. Likewise, SECURE~\cite{bhusal2024secure} tests knowledge in cybersecurity advisory through 2,036 questions based on MITRE ATT\&CK and CWE. Despite these efforts, existing studies only assess LLMs based on fragmented knowledge and lack a comprehensive model of the knowledge and skills needed by a cybersecurity expert. Consequently, these assessments do not address the questions posed in this paper. To bridge these gaps, this paper introduces a comprehensive assessment framework involving 345 knowledge points across 7 subdomains, with 11,050 high-quality questions specifically designed to evaluate LLMs' cybersecurity capabilities.
\section{CSEBenchmark}
\label{sec:method}

This paper introduces a cognitive science-based cybersecurity expert knowledge framework, which forms the foundation of CESBenchmark, the first evaluation dataset designed to assess the capabilities of LLMs in progressing toward a digital cybersecurity expert. The construction process is shown in Figure~\ref{fig:overview}, which is divided into four steps: developing the knowledge framework (Section~\ref{subsec:method_knowledge_framework}), classifying the knowledge points (Section~\ref{subsec:method_knowledge_classification}), collecting targeted materials and generating questions based on the classified knowledge points (Section~\ref{subsec:method_question_generation}), and validating and correcting the generated questions (Section~\ref{subsec:method_dataset_validation}).

\subsection{Knowledge Framework}
\label{subsec:method_knowledge_framework}

To evaluate whether LLMs can function as digital cybersecurity experts, we need to assess whether they possess the knowledge that a human cybersecurity expert should have, which is often documented in roadmaps. A roadmap is a structured guide that outlines the essential skills and knowledge required for a particular role. In this study, we select the well-known community-driven roadmap website, \textit{roadmap.sh}, as our source. This project has gained 295k stars on GitHub and provides a detailed overview of the skills and knowledge needed for various roles in the IT industry.  We use the \textit{Cybersecurity Expert Roadmap}~\cite{cybersecurity_expert_roadmap} and the \textit{Ethical Hacking Roadmap}~\cite{ethical_hacking_roadmap} as the basis for the CSEBenchmark knowledge framework. Additionally, we supplement our framework with the roadmap titled \textit{``From Power Button to PWN: A Roadmap to Computer Security,''}~\cite{computer_security_roadmap} collected from \textit{Hacking \& Coding Discord} communities.

Based on these three roadmaps, we develop a cybersecurity expert knowledge framework, as illustrated in Listing~\ref{lst:framework_example}. This framework consists of seven subdomains, each representing a key area of expertise for cybersecurity professionals: Fundamental IT Skills (FIS), Operating Systems (OS), Networking Knowledge (NK), Web Knowledge (WK), Security Skills and Knowledge (SSK), Cloud Skills and Knowledge (CSK), and Programming Skills and Knowledge (PSK). Each subdomain is organized into a hierarchical tree structure, with knowledge points arranged by level, culminating in 345 leaf nodes that represent the most specific knowledge points. This structure enables a fine-grained assessment of cybersecurity experts, offering a comprehensive depiction of the core knowledge required in the field.

\subsection{Knowledge Classification}
\label{subsec:method_knowledge_classification}

As discussed previously, different types of knowledge require varying levels of mastery. Cybersecurity, as an interdisciplinary field, spans both theoretical and practical domains. It encompasses knowledge points that include factual content to be memorized, concepts that require deep understanding, and skills that require hands-on practice. This framework aligns well with the cognitive science knowledge
\begin{lstlisting}[language=json, label={lst:framework_example}, caption={Example of the knowledge framework.}]
{"Cyber Security": {
    "Fundamental IT Skills": {
        "Common computer formats": {
            "label": "factual"
        }, ...
    },
    "Operating Systems": {
        "Windows": {
            "User management in Windows": {
                "label": "conceptual"
            }, ...        
        }, ...
    },
    "Networking Knowledge": {
        "Understand Common Protocols": {
            "TCP": {
                "label": "conceptual"
            }, ...
        },...
    },
    "Web Knowledge": {
        "SQL": {
            "label": "procedural",
        }, ...
    },
    "Security Skills and Knowledge": {
        "Footprinting and Reconnaissance": {
            "Google Dorks": {
                "label": "procedural"
            }, ...
        }, ...
    },
    "Cloud Skills and Knowledge": {
        "IaaS": {
            "label": "conceptual"
        }, ...
    },
    "Programming Skills and Knowledge": {
        "Python": {
            "label": "procedural"
        }, ...
    }
}
\end{lstlisting}
\noindent classification theory~\cite{bloom1956taxonomy}, which serves as the basis for categorizing cybersecurity knowledge in this study into factual, conceptual, and procedural types. These categories correspond to specific information, theoretical understanding, and practical skills, respectively. This classification enables a more nuanced evaluation of knowledge mastery, allowing an accurate and tailored assessment of each knowledge point.

To classify the 345 knowledge points in the CSEbenchmark knowledge framework, we invite two cybersecurity practitioners to label each point based on their understanding of the required level of mastery. When disagreements arise, a more experienced cybersecurity expert is consulted for a final decision. This process results in 121 factual knowledge points, 136 conceptual knowledge points, and 88 procedural knowledge points, with examples shown in Listing~\ref{lst:framework_example}. These labels reflect practitioners' views on the necessary level of understanding for each knowledge point, making the CSEBenchmark more aligned with real-world practices.

\subsection{Question Generation}
\label{subsec:method_question_generation}

After completing the knowledge classification, it is essential to generate targeted questions suited to each type of knowledge. First, we need to collect targeted material: for factual knowledge, brief descriptions from the roadmap or relevant wiki entries serve as primary sources for question generation, as factual knowledge mainly requires recall, and these sources provide direct, relevant content. For conceptual knowledge, we select insights from reputable websites or content sourced from textbooks, as these materials often include the author's understanding of the knowledge points, which help assess the test subject's deeper understanding of the concepts. For procedural knowledge, official documentation or tutorials are referenced, since they outline practical steps, meeting the needs for evaluating proficiency in hands-on tasks. Following these criteria, we manually collected the most relevant English material entry for each knowledge point to support effective question generation. We use the \textit{pymupdf4llm}~\cite{pymupdf4llm} library to convert PDFs to markdown format and manually preprocess the material to remove irrelevant text, such as image references, while restoring the original chapter structure information for use in subsequent steps.

After collecting targeted materials, we utilize an LLM to automatically generate questions from them, producing one correct answer and three distractors for each question. Specifically, we use the GPT-4-turbo model for question generation, given its strong performance in text processing. To help the model accurately grasp the characteristics of different knowledge types, we first define the question for each knowledge category in the prompt (see Table~\ref{tab:question_definition}). These definitions clarify the focus of the questions across knowledge types, ensuring that the model accurately reflects the unique attributes of each type. To further guide the model, we provide eight human-generated sample questions for each knowledge type, helping it recognize the distinct characteristics of each category and avoid misclassification. When generating questions, we explicitly specify the relevant knowledge type and emphasize the exclusion of unrelated categories. The model then selects the correct answer from the provided material and generates three distractors, ensuring the questions meet our expectations.

\begin{table}[htbp]
    \centering
    \footnotesize
    \caption{Definitions for question generation across different knowledge types.}
    \label{tab:question_definition}
    \begin{tabular}{cl}
    \hline
    \textbf{Type} & \multicolumn{1}{c}{\textbf{Definition}} \\ \hline
    \multirow{2}{*}{Factual} & Multiple-choice questions focusing on factual know-\\
        & ledge emphasize memory and recall. \\ \hline
    \multirow{3}{*}{Conceptual} & Multiple-choice questions focusing on conceptual \\
        & knowledge emphasize understanding and applying \\
        & abstract concepts \\ \hline
    \multirow{5}{*}{Procedural} & Multiple-choice questions focusing on procedural \\
        & knowledge emphasize the mastery of specific opera-\\
        & tional steps and procedural skills, particularly in the \\
        & context of solving targeted problems within defined \\ 
        & scenarios. \\ \hline
    \end{tabular}
\end{table}

Due to the limited input window of the LLM, it cannot process all of the materials at once. Additionally, overly lengthy material may lead the model to overlook important details, necessitating the division of the material. The conventional approach involves setting a token threshold and splitting the material into smaller segments~\cite{tihanyi2024cybermetric}. However, this method may disrupt the structure of the material, resulting in a loss of contextual information. To avoid this issue, we divide the material according to its chapter structure, ensuring that each section retains complete contextual integrity after segmentation.

We observe that materials of the same length may differ in information density. For materials with a higher information density, a greater number of questions should be generated, while for those with lower information density, fewer questions are appropriate. An inappropriate number of questions could lead to repetition or inadequate coverage of the material. Therefore, we aim to quantify the information density of the material and adaptively determine the number of questions to generate. Specifically, we define information density as the number of topics, reframing the task of quantifying information density as a topic extraction problem---a task easily handled by the LLM. In the prompt, we instruct the LLM to first identify all topics and then generate five questions per topic, achieving an adaptive match between the number of questions and the information density of the material. 

We generate a total of 11,743 questions for 345 knowledge points. To eliminate the impact of duplicate questions, we apply Semantic Textual Similarity for deduplication. We use SentenceTransformers~\cite{reimers-2019-sentence-bert} to convert questions into vectors and apply a similarity threshold of 0.85, validated experimentally for accuracy, to identify and remove duplicates. When duplicates are detected, only the earlier occurrence is retained. Following the question generation process, we obtain a final set of 11,468 unique questions, incurring a total cost of \$234.5.

\subsection{Dataset Validation and Correction}
\label{subsec:method_dataset_validation}

Due to the well-known issue of hallucination~\cite{zhang2023siren}, questions generated by the LLM are not always reliable. To address this, we conduct manual validation and correction of the 11,468 deduplicated questions. Specifically, we engage human annotators with cybersecurity expertise to answer each question without access to the original material, avoiding the potential influence of any inaccuracies in the source content. When discrepancies arise between the \removedtext{actual}\newtext{expert responses} and \removedtext{generated}\newtext{LLM-produced} answers, a senior cybersecurity expert conducts a secondary review to ensure accuracy. The entire validation process takes a total of 672 man-hours.

During validation, we find that 1,726 questions exhibit the following issues: (1) 384 questions contain incorrect answers; (2) 298 questions have multiple correct options; (3) 261 questions lack context in the question stem, resulting in incomplete or hard-to-understand questions; (4) 7 questions display a mismatch in question type; (5) 397 questions show weak relevance to the knowledge point; (6) 216 questions have low-quality distractors that are overly simple or obvious; (7) 14 questions are duplicates of other questions, despite having passed initial similarity checks; and (8) 149 questions lack a correct option. We attempt to manually correct these problematic questions. For issue (1), we replace the incorrect answer directly. For issues (2) and (6), we use the LLM to generate three similar but incorrect options based on the correct answer. For issues (3) and (8), we replace the correct answer or add the missing context based on annotators' feedback. For issues (4), (5), and (7), we remove these questions as they do not contribute to an accurate assessment. In total, we successfully corrected 1,308 problematic questions, enhancing the CSEBenchmark dataset.

\begin{table}[htbp]
    \centering
    \footnotesize
    \caption{Distribution of Knowledge Points and Questions Across Subdomains in the CSEBenchmark Dataset.}
    \label{tab:distribution}
    \begin{tabular}{ccccc}
    \hline
       \textbf{Subdomain} & \textbf{Type} & \textbf{\#Knowledge} & \newtext{\textbf{\#Tokens}} & \textbf{\#Questions} \\\hline
        \multirow{3}{*}{FIS} & Factual & 21 & \newtext{19.8K} & 124 \\
            & Conceptual & 2 & \newtext{3.3K} & 12 \\
            & Procedural & 2 & \newtext{18.7K} & 25 \\ \hline
        \multirow{3}{*}{OS} & Factual & 5 & \newtext{8.4K} & 25 \\
            & Conceptual & 18 & \newtext{0.3M} & 433 \\
            & Procedural & 16 & \newtext{0.4M} & 650 \\ \hline
        \multirow{3}{*}{NK} & Factual & 30 & \newtext{14.9K} & 168 \\
            & Conceptual & 31 & \newtext{0.6M} & 757 \\
            & Procedural & 12 & \newtext{93.2K} & 140 \\ \hline
        \multirow{3}{*}{WK} & Factual & 0 & \newtext{0} & 0 \\
            & Conceptual & 0 & \newtext{0} & 0 \\
            & Procedural & 6 & \newtext{1.8M} & 2202 \\ \hline
        \multirow{3}{*}{SSK} & Factual & 50 & \newtext{22.2K} & 268 \\
            & Conceptual & 79 & \newtext{0.9M} & 1040 \\
            & Procedural & 46 & \newtext{2.0M} & 2451 \\ \hline
        \multirow{3}{*}{CSK} & Factual & 15 & \newtext{15.7K} & 75 \\
            & Conceptual & 6 & \newtext{91.3K} & 144 \\
            & Procedural & 0 & \newtext{0} & 0 \\ \hline
        \multirow{3}{*}{PSK} & Factual & 0 & \newtext{0} & 0 \\
            & Conceptual & 0 & \newtext{0} & 0 \\
            & Procedural & 6 & \newtext{2.0M} & 2536 \\ \hline
        \textbf{Count} & & 345 & \newtext{8.4M} & 11,050 \\\hline
    \end{tabular}
\end{table}

The finalized CSEBenchmark dataset comprises 11,050 high-quality multiple-choice questions, covering seven subdomains. The distribution of question types and quantities is shown in Table~\ref{tab:distribution}. \newtext{Notably, the distribution of knowledge points and questions exhibits a skew, primarily driven by two factors: inherent variations in knowledge point distribution across subdomains, which stem from the roadmap design, and the uneven distribution of questions, which correlates with the token count in each corpus, as larger corpus naturally encompass a greater number of topics.}

\section{Experimental Investigation}
\label{sec:evaluation}

\subsection{Experiment Settings}
\label{subsec:experiment_settings}

\noindent\textbf{LLM selection and configuration.} In this study, we select \removedtext{10}\newtext{12} state-of-the-art LLMs for evaluation, as shown in Table~\ref{tab:models}. These models have demonstrated strong performance in text processing and are widely applied across various tasks. The selected models include both popular open-source models and several commercial closed-source models, with parameter scales ranging from 3B to \removedtext{175B}\newtext{671B}, reflecting the cybersecurity knowledge capabilities of models at different scales. Specially, we introduce a mixture-of-experts (MoE) model, Mixtral 8$\times$7B, which consists of 8 experts, each with 7B parameters, totaling approximately 45B parameters. \newtext{We also introduce an inference model, Deepseek-R1, which is trained on Deepseek-V3 and, unlike other models, autonomously generates its own chain of thought, systematically deducing intermediate steps to ensure accurate reasoning and logical coherence.} For OpenAI \newtext{and Deepseek} models, we access them via their respective APIs~\cite{openaiapireference, deepseekapireference}, while for other open-source models, we use the OpenAI-Compatible Server from \textit{vLLM}~\cite{kwon2023efficient} to ensure code consistency. To assess the knowledge levels of these models more precisely, we set the temperature parameter to 0.2, which is commonly used in precision tasks~\cite{openaiapireference}, to minimize the influence of random output on evaluation results.

\begin{table}[htbp]
    \centering
    \footnotesize
    \caption{Selected LLMs in this study.}
    \label{tab:models}
    \begin{tabular}{cccc}
    \hline
       \textbf{Model Name} & \textbf{\#Params} & \textbf{Cutoff Date} & \textbf{Type} \\ \hline
       GPT-3.5-Turbo-0125 & 175B & 2021-09 & Closed\\
       GPT-4-Turbo-2024-04-09 & Unk. & 2023-12 & Closed\\
       GPT-4o-2024-08-06 & Unk. & 2023-10 & Closed\\
       Llama-3.2-3B-Instruct & 3B & 2023-12 & Open\\
       Llama-3.1-8B-Instruct & 8B & 2023-12 & Open\\
       Llama-3.1-70B-Instruct & 70B & 2023-12 & Open\\
       Mixtral-8x7B-Instruct-v0.1 & 45B & 2023-12 & Open\\
       Qwen-2.5-3B-Instruct & 3B & 2023-02 & Open\\
       Qwen-2.5-7B-Instruct & 7B & 2023-02 & Open\\
       Qwen-2.5-72B-Instruct & 72B & 2023-02 & Open\\
       \newtext{Deepseek-V3-241226} & \newtext{671B} & \newtext{Unk.} & \newtext{Open} \\
       \newtext{Deepseek-R1-250120} & \newtext{671B} & \newtext{Unk.} & \newtext{Open} \\ \hline
    \end{tabular}
\end{table}

\noindent\textbf{Platform.} The experiments are conducted on a platform with an Intel(R) Xeon(R) Platinum 8468 processor, 2.0 TB RAM, 172 cores and 8 NVIDIA H100 GPUs with 80 GB HBM3 each. The entire experiment requires a total of 1.08 GPU-weeks.

\noindent\textbf{Experiment Setup.} Recognizing that different prompts can influence how the models activate their embedded knowledge, we employ three interaction methods---Zero-shot, Few-shot, and CoT---in our experiments to minimize the impact of these prompting techniques on the models' output~\footnote{\newtext{For Deepseek-R1, since it inherently incorporates the CoT method, we only use the CoT approach.}}. For each question, we use the highest score from the three prompting methods as the final result, representing the actual knowledge ceiling that the model can achieve. In the Zero-shot method, we provide questions directly without any examples, asking the model to produce results independently. For the Few-shot method, we build on the Zero-shot approach by providing 5 example question-answer pairs that are not included in the dataset; this 5-shot strategy is widely used in related research~\cite{liu2023secqa, liu2024cyberbench}. Finally, in the CoT method, we use the common prompt, \textit{``Let's think step by step,''} to guide the model's reasoning process. Full prompts are provided in the Appendix~\ref{appendix:prompt_answer_generation}.

\noindent\textbf{Measurement Method.} To reduce the impact of LLM randomness on the evaluation results, we have the model perform five independent inferences for each question, considering the response correct only if all of the inferences yield the correct answer. Additionally, to avoid any preference the model may have for specific options, we systematically rotate the correct answer across the four choices (A, B, C, D) and evaluate each arrangement independently. We consider the model to have truly mastered a knowledge point only if it answers correctly in all four arrangements, indicating that its success is due to understanding rather than guessing.

Given that LLM outputs are in the loose format of natural language text, we need to extract the exact options selected by the models. A common approach is to evaluate the probability of the first token in the model output~\cite{hendryckstest2021, hendrycks2021ethics}; however, recent research indicates that this method lacks robustness~\cite{wang2024my, wang2024look}. Therefore, we follow their recommendations to extract the model's selected answers from the original responses. Specifically, we use the xFinder-llama38it model for option extraction, a state-of-the-art model for identifying multiple-choice answers, which has demonstrated 95.47\% accuracy on generalization sets~\cite{xFinder}. We randomly sample 4782 original responses for manual verification, finding an actual accuracy of 92.47\% for this extraction process, which supports the validity of the results presented in this study.

\noindent\textbf{Evaluation Metrics.} We use the accuracy for all questions associated with each knowledge point as our evaluation metric, categorizing accuracy into four ranges: 100\% indicates that the LLMs have fully mastered the knowledge point, meeting the level expected of cybersecurity experts; \([90\%, 100\%)\) suggests that LLMs are approaching expert-level understanding; \([80\%, 90\%)\) indicates partial mastery with room for improvement; and below 80\% reflects poor performance, indicating areas that require focused attention.

\noindent\textbf{Research Question.} In the following subsections, we evaluate the performance of the selected \removedtext{10}\newtext{12} state-of-the-art LLMs in the CSEBenchmark, with a primary focus on the following research questions:

\noindent\textbf{RQ1.} Do the selected LLMs possess the knowledge expected of cybersecurity experts?

\noindent\textbf{RQ2.} What knowledge gaps remain in the selected LLMs when positioned as cybersecurity experts?

\noindent\textbf{RQ3.} Can the results of CSEBenchmark help improve LLM performance in cybersecurity tasks?

\noindent\textbf{RQ4.} How well do the selected LLMs align with real-world cybersecurity job roles?

\begin{table*}[ht]
    \centering
    \footnotesize
    \caption{Accuracy of the tested LLMs across seven subdomains and three knowledge categories (acronyms used).}
    \label{tab:results}
    \begin{tabular}{cccccccccccccc}
    \hline
        \multirow{2}{*}{\textbf{Type}} & \multirow{2}{*}{\textbf{Label}} & \textbf{GPT-} & \textbf{GPT-} & \textbf{GPT-} & \textbf{L3.1-} & \textbf{L3.1-} & \textbf{L3.2-} & \textbf{M-} & \textbf{Q2.5-} & \textbf{Q2.5-} & \textbf{Q2.5-} & \newtext{\textbf{DS-}} & \newtext{\textbf{DS-}}\\
            & & \textbf{3.5T} & \textbf{4T} & \textbf{4o} & \textbf{8B} & \textbf{70B} & \textbf{3B} & \textbf{8$\times$7B} & \textbf{3B} & \textbf{7B} & \textbf{72B} & \newtext{\textbf{V3}} & \newtext{\textbf{R1}}\\\hline
        \multirow{7}{*}{Subdomain} & FIS & 87.58 & 92.55 & 95.65 & 88.20 & 91.30 & 80.75 & 86.34 & 87.58 & 91.30 & \textbf{96.27} & \newtext{93.79} & \newtext{91.93}\\
            & OS & 61.91 & 80.60 & \textbf{82.67} & 64.08 & 74.37 & 48.83 & 69.95 & 65.25 & 69.58 & 80.60 & \newtext{81.32} & \newtext{79.87}\\
            & NK & 81.03 & 91.46 & 92.39 & 83.19 & 88.64 & 70.61 & 84.32 & 79.72 & 87.23 & \textbf{92.58} & \newtext{91.92} & \newtext{89.86}\\
            & WK & 67.94 & 84.74 & \textbf{86.15} & 67.71 & 80.79 & 49.41 & 72.48 & 64.80 & 72.93 & 84.11 & \newtext{84.65} & \newtext{79.16}\\
            & SSK & 62.92 & 78.21 & \textbf{80.26} & 65.28 & 74.57 & 51.74 & 68.79 & 65.79 & 70.44 & 79.76 & \newtext{79.70} & \newtext{74.79}\\
            & CSK & 87.67 & 95.89 & 97.26 & 92.24 & 95.43 & 83.56 & 92.24 & 88.13 & 93.61 & 96.35 & \newtext{\textbf{97.72}} & \newtext{96.80}\\
            & PSK & 71.77 & 88.13 & 89.04 & 69.91 & 84.15 & 47.79 & 76.30 & 67.67 & 77.68 & 87.97 & \newtext{\textbf{89.87}} & \newtext{85.29}\\ \hline
        \multirow{3}{*}{Category} & Fact. & 86.82 & 93.64 & \textbf{94.85} & 86.06 & 92.42 & 80.00 & 88.33 & 87.58 & 90.45 & 94.24 & \newtext{94.24} & \newtext{91.67}\\
            & Conc. & 86.25 & 93.88 & \textbf{94.84} & 88.60 & 93.34 & 78.54 & 89.52 & 86.34 & 91.32 & 94.59 & \newtext{94.26} & \newtext{92.58}\\
            & Proc. & 61.62 & 80.07 & \textbf{81.83} & 62.17 & 75.00 & 43.09 & 67.62 & 61.02 & 68.72 & 80.55 & \newtext{81.37} & \newtext{76.14}\\ \hline
        Overall & & 68.44 & 83.86 & \textbf{85.42} & 69.30 & 80.00 & 52.95 & 73.58 & 68.07 & 74.90 & 84.40 & \newtext{84.92} & \newtext{80.62}\\ \hline
    \end{tabular}
\end{table*}

\subsection{LLM Cybersecurity Expertise Assessment (RQ1)}
\label{subsec:experiment_knowledge_assessment}

Table~\ref{tab:results} presents the accuracy performance of the 12 selected LLMs on CSEBenchmark. Overall, GPT-4o ranks first with an accuracy of 85.42\%, \removedtext{followed by Qwen-2.5-72B with 84.40\%. In third place is GPT-4-Turbo, with an accuracy of 83.86\%, trailing the top two models by less than 1.8\%.}\newtext{followed closely by Deepseek-V3 at 84.92\% and Qwen-2.5-72B at 84.40\%, with less than a 1.2\% difference among the top three models. GPT-4-Turbo follows in fourth place at 83.86\%~\footnote{Note that since GPT-4-Turbo is also used for generating the questions, its results may involve cyclical use, as discussed in Section~\ref{subsec:bias}.}. Deepseek-R1 and Llama-3.1-70B achieve 80.62\% and 80.00\%, respectively.} The remaining models show a larger performance gap of over 5\% compared to the top \removedtext{three}\newtext{six}, with the rankings as follows: \removedtext{Llama-3.1-70B (80.00\%), }Qwen-2.5-7B (74.90\%), Mixtral-8$\times$7B (73.58\%), GPT-3.5-Turbo (68.44\%), Llama-3.1-8B (69.30\%), Qwen-2.5-3B (68.07\%), and Llama-3.2-3B (52.95\%). Notably, GPT-4o not only performs well in terms of accuracy but also operates at just 30\% of the cost of GPT-4-Turbo, making it a preferred choice among closed-source LLMs for cybersecurity expert scenarios. Among open-source LLMs, \removedtext{Qwen-2.5-72B}\newtext{Deepseek-V3} performs the best, coming close to the top-performing GPT-4o. Due to its open-source nature, \removedtext{Qwen-2.5-72B}\newtext{Deepseek-V3} also offers greater scalability and practicality. \newtext{Notably, although Qwen-2.5-72B's accuracy is slightly lower than Deepseek-V3 (0.6\%), its substantially smaller model size (72B vs. 671B) makes it a more cost-effective and practical choice for real-world applications.} We observe that the Qwen-2.5 series consistently outperforms the Llama-3.1 and Llama-3.2 series of similar parameter scales. Additionally, the Mixtral-8$\times$7B MoE model lags behind the single 7B model, Qwen-2.5-7B. Although the MoE structure is theoretically designed to enhance performance through specialized expert modules, it does not show a significant advantage in this evaluation, suggesting that the multi-expert mechanism has limited effectiveness for knowledge tasks in this context. \newtext{We also observe that, despite Deepseek-R1's strong reasoning capabilities, it does not exhibit an advantage in the safety knowledge evaluation. Its overall accuracy is even lower than that of its training base, Deepseek-V3. This suggests that in knowledge tasks, strong reasoning ability may not necessarily compensate for precise knowledge recall and retrieval. Over-reliance on reasoning could instead lead to information distortion or misjudgment.}

\begin{findingbox}
    GPT-4o is the best-performing LLM overall, while \removedtext{Qwen-2.5-72B}\newtext{Deepseek-V3} leads among open-source options. However, even these top LLMs cover only 85.42\% of the knowledge required by cybersecurity experts.
\end{findingbox}

\begin{figure}[htbp]
    \centering
    \includegraphics[width=0.7\linewidth]{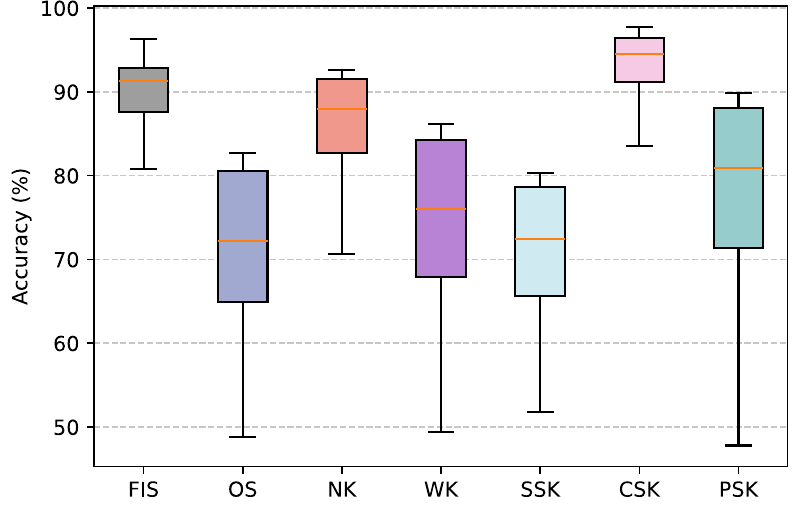}
    \caption{Accuracy distribution of LLMs across subdomains.}
    \label{fig:subdomain_box}
\end{figure}

In all subdomains, GPT-4o performs best in \removedtext{five}\newtext{three}—OS (82.67\%), WK (86.15\%), SSK (80.26\%), CSK (97.26\%), and PSK (89.04\%)—while \newtext{Deepseek-V3 leads in CSK (97.72\%) and PSK (89.87\%), and} Qwen-2.5-72B leads in FIS (96.27\%) and NK (92.58\%). Although current LLMs do not fully meet the knowledge requirements of security experts, their highest accuracies exceed 90\% in the FIS, NK, and CSK subdomains, indicating that their knowledge in these areas is approaching cybersecurity experts. Figure~\ref{fig:subdomain_box} presents a box plot of LLM accuracy across each subdomain. In the FIS and CSK subdomains, all LLMs achieve accuracies above 80\%, with a median of \removedtext{90}\newtext{91}\%, indicating that the tested LLMs are generally approaching the knowledge level of cybersecurity experts in these areas. In the NK subdomain, LLM performance varies widely, with the lowest accuracy at 71\% and a median of \removedtext{86}\newtext{88}\%. Although the top-performing LLMs exceed 90\% accuracy in this subdomain, most LLMs still have substantial room for improvement in knowledge coverage. In the OS, WK, SSK, and PSK subdomains, accuracy differences among LLMs increase significantly, with the lowest accuracy falling below 51\% and a median slightly above \removedtext{69}\newtext{72}\%, indicating lower knowledge levels in these subdomains.

\begin{findingbox}
    LLMs have not yet fully met the knowledge requirements of cybersecurity experts in any subdomain. However, their knowledge in the FIS, NK, and CSK subdomains is close to the expert level, while significant improvement is needed in the OS, WK, SSK, and PSK subdomains.
\end{findingbox}

\begin{figure}[htbp]
    \centering
    \includegraphics[width=0.7\linewidth]{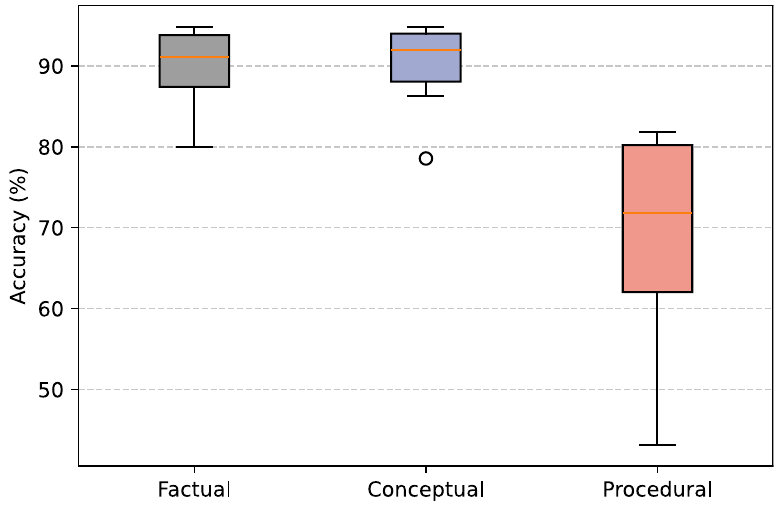}
    \caption{Accuracy distribution of LLMs across knowledge categories.}
    \label{fig:category_box}
\end{figure}

We also evaluate the accuracy of the tested LLMs across three knowledge categories, with results presented in Table~\ref{tab:results}. GPT-4o achieves the highest accuracy across all three categories, at 94.85\%, 94.84\%, and 81.83\%, respectively, with the ranking of the other models remaining largely consistent with their overall performance. The accuracy distribution across these categories is illustrated in the box plot in Figure~\ref{fig:category_box}. In the Factual and Conceptual categories, the accuracy of LLMs is relatively concentrated, with almost all models achieving close to 80\% accuracy and a median close to \removedtext{90}\newtext{92}\%, indicating that LLMs are adept at mastering these types of knowledge. This may be because factual and conceptual knowledge often appears in direct statements or explanatory forms within the training corpus, allowing models to extract and retain information from context more effectively. In contrast, the accuracy drops significantly for procedural knowledge, with the lowest accuracy at only 43.09\% and a median of \removedtext{68.17}\newtext{71.86}\%. This discrepancy likely arises because LLM pretraining is not tailored to reinforce real-world cybersecurity operations or procedural tasks, making it challenging for models to develop a deep understanding and flexible application of complex operations from the corpus alone. Given that cybersecurity heavily relies on practical skills, this limitation presents a significant obstacle for LLMs to become cybersecurity experts.

\begin{findingbox}
    LLMs demonstrate a good grasp of factual and conceptual knowledge, but perform poorly in procedural knowledge.
\end{findingbox}

\begin{figure*}[htbp]
    \includegraphics[width=0.95\textwidth]{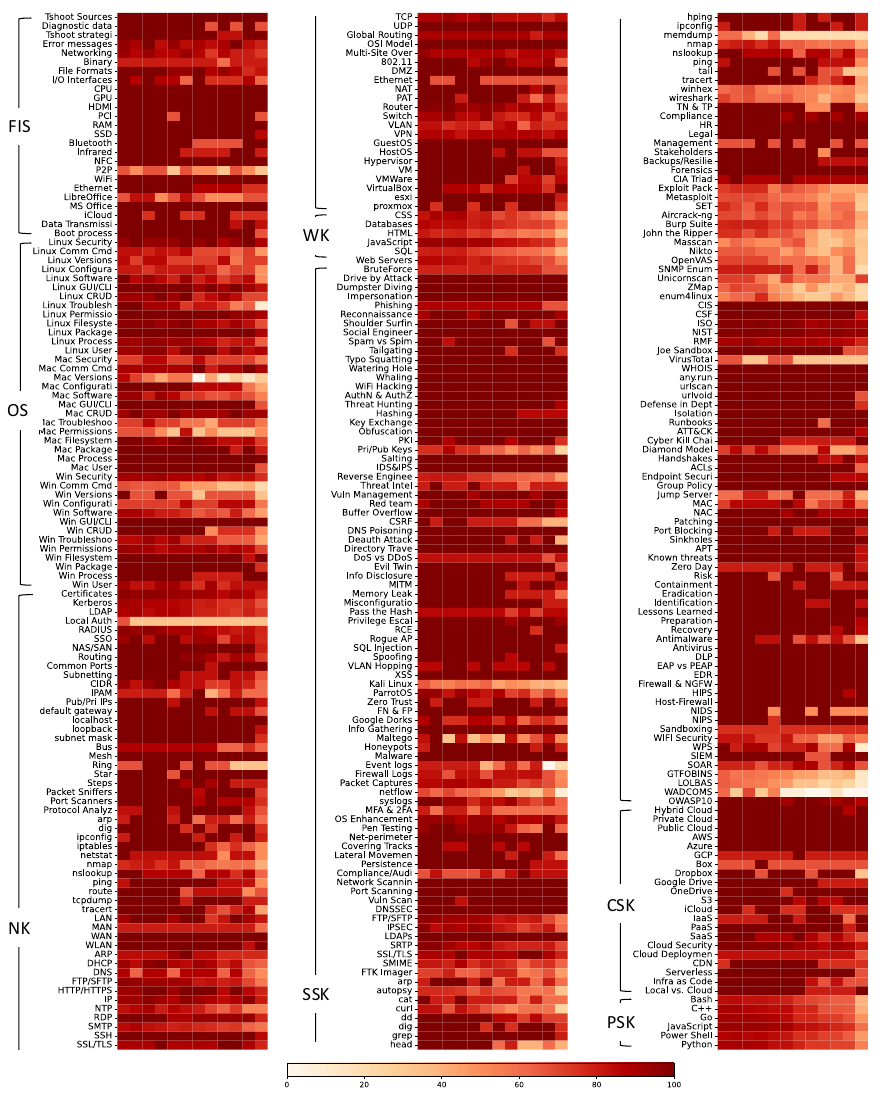}
    \caption{Heatmap of accuracy across 345 knowledge points for 12 models. The y-axis labels denote individual knowledge points, with subdomain names in parentheses for grouped items. Each section contains 12 columns representing models from left to right: GPT-4o, \newtext{Deepseek-V3, }Qwen-2.5-72B, GPT-4-Turbo, \newtext{Deepseek-R1, }Llama-3.1-70B, Qwen-2.5-7B, Mixtral-8x7B, GPT-3.5-Turbo, Llama-3.1-8B, Qwen-2.5-3B, Llama-3.2-3B.}
    \label{fig:heatmap}
\end{figure*}

We conduct a fine-grained evaluation of LLM performance across 345 knowledge points, with the results displayed as a heatmap in Figure~\ref{fig:heatmap}. In the heatmap, each row represents the accuracy of different LLMs on the same knowledge point, while each column shows the performance of the same LLM across various knowledge points. Among the 345 knowledge points, certain LLMs achieve 100\% accuracy on \removedtext{238}\newtext{241} points, indicating that LLMs meet the knowledge standards of security experts for these points. Additionally, on \removedtext{33}\newtext{35} knowledge points, certain LLMs reach an accuracy above 90\%, suggesting that LLMs are approaching expert-level knowledge in these areas. Of these \removedtext{271}\newtext{276} knowledge points, \removedtext{228}\newtext{230} are factual or conceptual knowledge, accounting for \removedtext{83.96}\newtext{83.33}\%, further confirming the strong performance of LLMs in these knowledge types. The remaining \removedtext{43}\newtext{46} knowledge points are procedural, focusing on essential operations and troubleshooting for operating systems and network tools. These include troubleshooting strategies, error interpretation, software installation on Linux, MacOS, and Windows, basic commands (e.g., \textit{ping}, \textit{netstat}), log analysis, file manipulation (e.g., \textit{cat}, \textit{grep}), and scripting languages (e.g., \textit{Python}, \textit{JavaScript}). Although these procedural knowledge points involve a degree of practical skill, their high frequency in real-world tasks means their fixed syntax and relatively simple logic are well-represented in pretraining data, enabling LLMs to achieve high accuracy on these points.

\begin{findingbox}
    LLMs achieve the expected level of cybersecurity expertise on \removedtext{238}\newtext{241} knowledge points and approach expert-level performance on an additional \removedtext{33}\newtext{35} points, covering \removedtext{78.5}\newtext{80.0}\% of all points. These are primarily factual and conceptual knowledge, along with some high-frequency procedural knowledge.
\end{findingbox}

\subsection{LLM Knowledge Gap Assessment (RQ2)}
\label{subsec:experiment_knowledge_gap}

As mentioned above, LLMs meet or approach the knowledge requirements of cybersecurity experts on \removedtext{271}\newtext{276} knowledge points, but notable knowledge gaps remain on the other \removedtext{74}\newtext{69} points. Benefiting from the fine-grained design of knowledge points in CSEBenchmark, we are able to analyze these specific knowledge gaps in each LLM in greater detail than existing studies that rely solely on overall score evaluations. Among these \removedtext{74}\newtext{69} knowledge points, \removedtext{45}\newtext{40} have accuracies between 80\% and 90\%, indicating that LLMs have a partial grasp of these points but still have room for improvement. Of these, \removedtext{12}\newtext{11} are factual knowledge points, covering topics like basic coding, operating system version differences, threat intelligence, authentication methods, and security models. Another \removedtext{12}\newtext{11} are conceptual knowledge points, addressing core security concepts and network protocols, such as \textit{MacOS permissions management}, \textit{DNS}, \textit{VPNs}, and \textit{DDoS attacks}. The remaining \removedtext{21}\newtext{18} points are procedural knowledge, primarily involving system operations, common commands, and tool applications, such as installation and configuration in Linux and Windows, network scanning tools (e.g., \textit{nmap}), log analysis (e.g., \textit{event logs}, \textit{packet captures}), introductory reverse engineering, and scripting and programming languages (e.g., \textit{Bash}, \removedtext{\textit{C++}, }\textit{PowerShell}).

\begin{figure}[htbp]
    \centering
    \includegraphics[width=0.8\linewidth]{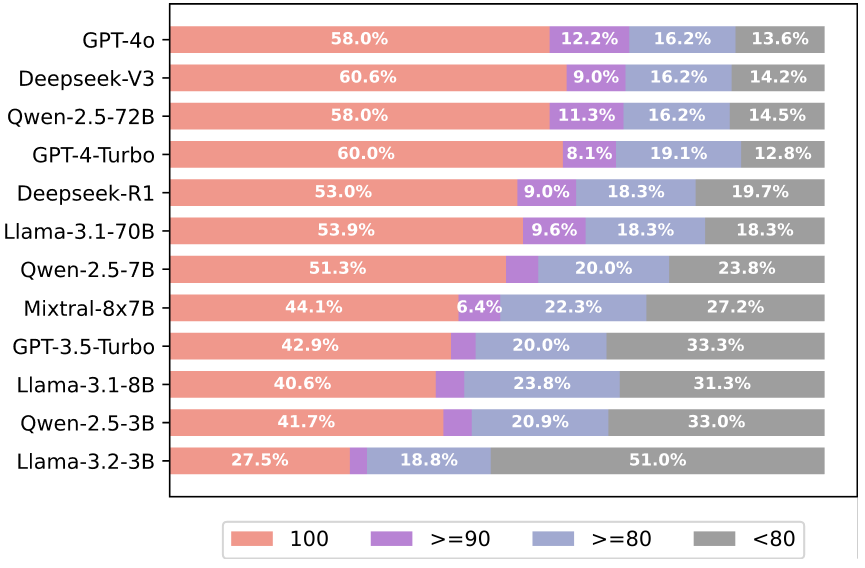}
    \caption{Proportion of knowledge points across four accuracy ranges for each LLM.}
    \label{fig:knowledge_bar}
\end{figure}

There are 29 knowledge points where the highest accuracy achieved by any LLM remains below 80\%, indicating substantial room for improvement in these areas. Of these, 4 are factual knowledge (\textit{P2P}, \textit{Local Auth}, \textit{VirusTotal}, and \textit{Sandboxing}) and 1 is conceptual knowledge (\textit{Brute Force vs Password Spray}). We observe that, although these points appear straightforward, LLMs still struggle with them. For instance, one question on \textit{Local Auth} is: \textit{``What additional security measure is recommended to enhance the security of a system using local authentication? A. Use of SSL B. Centralized user management C. Cloud-based authentication D. Reduction of password strength.''} The correct answer is A. However, when the position of the correct answer is shuffled with other options, LLMs often select the wrong answer, indicating that the model's understanding of this knowledge point is not solid. The remaining 24 points are procedural knowledge, involving the use of cybersecurity and forensic tools, including common Windows commands, SQL, Kali Linux, network analysis tools (e.g., \textit{netflow}, \textit{Wireshark}), forensic tools (e.g., \textit{FTK Imager}, \textit{Autopsy}, \textit{memdump}, \textit{winhex}), exploitation frameworks (e.g., \textit{Exploit Pack}, \textit{Metasploit}), social engineering tools (e.g., \textit{Social-Engineer Toolkit}), wireless security tools (e.g., \textit{Aircrack-ng}), penetration testing tools (e.g., \textit{Burp Suite}, \textit{John the Ripper}, \textit{Nikto}, \textit{OpenVAS}), system information gathering tools (e.g., \textit{enum4linux}), and malicious command libraries (e.g., \textit{GTFOBINS}, \textit{LOLBAS}, \textit{WADCOMS}). Compared to more commonly encountered tools mentioned above(e.g., \textit{cat} and \textit{grep}), these points are more specialized and have unique application contexts, resulting in lower representation in pretraining corpora and making it challenging for LLMs to effectively learn and master them.

\begin{findingbox}
    Overall, LLMs show notable gaps in nuanced procedural knowledge involving specialized tools and uncommon commands, even struggling with certain straightforward factual and conceptual points.
\end{findingbox}

We further analyze the knowledge gaps in each LLM, with the accuracy distribution across all knowledge points shown in Figure~\ref{fig:knowledge_bar}.

\noindent\textbf{GPT-4o:} As the best-performing LLM overall, GPT-4o achieves 100\% accuracy on 200 knowledge points and exceeds 90\% accuracy on an additional 42 points, covering 70.14\% of all knowledge points. However, its accuracy falls below 80\% on 47 points, primarily in areas such as foundational concepts (e.g., \textit{Peer-to-Peer (P2P)}, \textit{Private vs Public Keys}), security tool usage (e.g., \textit{VirusTotal}, \textit{Wireshark}, \textit{Metasploit}), attack and defense techniques (e.g., \textit{Brute Force vs Password Spray}), system configuration tasks (e.g., \textit{Common Commands in Windows}), and security management (e.g., \textit{SOAR}).

\newtext{\noindent\textbf{Deepseek-V3}: Deepseek-V3 is the best-performing open-source LLM, achieving 100\% accuracy on 209 knowledge points and exceeding 90\% accuracy on an additional 31 points, covering 69.57\% of all knowledge points. However, the model struggles with 49 knowledge points, particularly system and network fundamentals (e.g., \textit{P2P}, \textit{DNS}), authentication (e.g., \textit{MFA}, \textit{Jump Server}), security tools (e.g., \textit{VirusTotal}, \textit{Metasploit}), penetration testing (e.g., \textit{Burp Suite}, \textit{OpenVAS}), system configuration (e.g., \textit{Windows commands}, \textit{MacOS troubleshooting}), and data analysis (e.g., \textit{SQL}, \textit{NetFlow}). It also fails to differentiate system versions and privilege escalation techniques (e.g., \textit{GTFOBins}, \textit{LOLBAS}), highlighting gaps in practical cybersecurity knowledge.}

\noindent\textbf{Qwen-2.5-72B:} Qwen-2.5-72B \removedtext{is the best-performing open-source LLM}\newtext{also demonstrates strong performance}, achieving 100\% accuracy on 200 knowledge points and exceeding 90\% accuracy on  39 more, covering 69.28\% of all knowledge points. However, it struggles with 50 knowledge points, primarily in the following areas: foundational system and network concepts (e.g., \textit{Peer-to-Peer (P2P)}, \textit{iCloud}), security compliance and management (e.g., \textit{Roles of Compliance and Auditors}), security tool usage (e.g., \textit{VirusTotal}, \textit{Metasploit}), attack and defense techniques (e.g., \textit{Brute Force vs Password Spray}), system configuration tasks (e.g., \textit{Common Commands in Linux}), and basic programming and data query tools (e.g., \textit{SQL}, \textit{Google Dorks}).

\noindent\textbf{GPT-4-Turbo:} Ranking third overall, GPT-4-Turbo covers the most knowledge points with 100\% accuracy, achieving perfect scores on 207 points, and over 90\% accuracy on an additional 28 points, totaling 68.12\% of all knowledge points. However, the model's accuracy falls below 80\% on 44 points, mainly in the following areas: foundational system and access management concepts (e.g., \textit{Peer-to-Peer (P2P)}, \textit{Local Auth}), roles in security compliance and management, cryptography and authentication mechanisms (e.g., \textit{WPA vs WPA2 vs WPA3 vs WEP}, \textit{Brute Force vs Password Spray}), security tool usage (e.g., \textit{VirusTotal}, \textit{Metasploit}), basic programming and data query tools (e.g., \textit{SQL}, \textit{Google Dorks}), and system configuration tasks (e.g., \textit{Common Commands in Windows}).

\newtext{\noindent\textbf{Deepseek-R1:} While Deepseek-R1 excels in reasoning, its performance in security knowledge assessment is less remarkable. It achieves 100\% accuracy on 183 knowledge points and exceeds 90\% on 31 more, covering 62.03\% of the total. However, it falls short on 68 knowledge points, particularly in authentication and access control (e.g., \textit{MFA \& 2FA}, \textit{Jump Server}), network security (e.g., \textit{NIDS}, \textit{VLAN}, \textit{DNS}), security tools (e.g., \textit{VirusTotal}, \textit{Metasploit}, \textit{Wireshark}), penetration testing (e.g., \textit{Aircrack-ng}, \textit{OpenVAS}, \textit{Masscan}), system administration (e.g., \textit{Linux installation}, \textit{Windows commands}), and forensic analysis (e.g., \textit{FTK Imager}, \textit{WinHex}).}

\noindent\textbf{Llama-3.1-70B:} Llama-3.1-70B shows a noticeable gap from the top \removedtext{three}\newtext{five} models, achieving 100\% accuracy on only 186 knowledge points, with an additional 33 points exceeding 80\% accuracy, covering 63.48\% of all knowledge points. The model performs poorly on 63 knowledge points, primarily in the following areas: operating system versions and configuration management (e.g., \textit{Different Versions and Differences in Linux}, \textit{Local Auth}), network interfaces and standards (e.g., \textit{Ethernet}, \textit{VLAN}), cloud storage and virtualization tools (e.g., \textit{iCloud}, \textit{VirtualBox}), security tool usage (e.g., \textit{VirusTotal}, \textit{Metasploit}), encryption and authentication mechanisms (e.g., \textit{Private vs Public Keys}, \textit{Brute Force vs Password Spray}), and basic programming operations (e.g., \textit{SQL}, \textit{Bash}).

\noindent\textbf{Qwen-2.5-7B:} As the best-performing small model, Qwen-2.5-7B achieves 100\% accuracy on 177 knowledge points, with an additional 17 points exceeding 90\%, covering 56.23\% of all knowledge points. However, the model's accuracy falls below 80\% on 82 points, particularly in areas such as network and communication protocols (e.g., \textit{Bluetooth}, \textit{Peer-to-Peer (P2P)}, \textit{Ethernet}), operating systems and file management (e.g., \textit{Linux version differences}, \textit{common Windows commands}, \textit{MacOS troubleshooting}), and authentication and security (e.g., \textit{MFA\&2FA}, \textit{user permissions management}, \textit{types of password attacks}). Additionally, the model shows weaker recognition and understanding in information gathering and vulnerability scanning tools (e.g., \textit{nmap}, \textit{Masscan}, \textit{Unicornscan}), data forensics and analysis tools (e.g., \textit{Wireshark}, \textit{FTK Imager}, \textit{Event Logs}), and foundational web and database knowledge (e.g., \textit{HTML}, \textit{SQL}, \textit{Web Servers}).

\noindent\textbf{Other five LLMs:} The remaining 5 LLMs have fewer knowledge points with 100\% accuracy or above 90\%, with coverage below 50\%, indicating that these LLMs fall short of the expected knowledge level for cybersecurity experts on more than half of the points. Among them, Mixtral-7$\times$8B has 27.8\% of knowledge points with accuracy below 80\%, while GPT-3.5-Turbo, Llama-3.1-8B, and Qwen-2.5-3B each have around 30\% of points below this threshold. Llama-3.2-3B performs the worst, with 51\% of knowledge points below 80\% accuracy. These results suggest that these models are currently insufficient for performing at a cybersecurity expert level.

Interestingly, we observe that different-sized LLMs within the same series also exhibit variations in their knowledge gaps. This suggests that the knowledge gaps of smaller LLMs are not merely a subset of those found in larger models. In fact, larger models may have gaps in areas where smaller models perform well. For example, Llama-3.1-70B underperforms when using \textit{tcpdump}, while Llama-3.1-8B achieves 100\% accuracy on this knowledge point. This highlights the importance of not relying solely on model size when selecting an LLM, but instead considering the specific tasks and knowledge gaps to make a more informed choice.

\begin{findingbox}
    Different LLMs exhibit distinct knowledge gaps as cybersecurity experts. Even smaller models in the same series can sometimes outperform larger ones in specific knowledge points.
\end{findingbox}

\subsection{Enhancing LLMs Through CSEBenchmark (RQ3)}
\label{subsec:experiment_enhance}

After identifying the knowledge gaps of each LLM using CSEBenchmark, we attempt to improve their performance based on these gaps. To this end, we \newtext{focus on two fundamental security tasks—vulnerability detection and threat intelligence analysis—and }select three state-of-the-art open-source, task-based evaluation datasets---VuldetectBench~\cite{liu2024vuldetectbench}, SecLLMHolmes~\cite{ullah2024llms}, and CTI-RCM~\cite{alam2024ctibench}\removedtext{---which represent LLM capabilities in vulnerability existence detection and threat intelligence analysis}. \newtext{VuldetectBench and SecLLMHolmes focus on vulnerability detection, with the former containing 1,000 real-world vulnerability snippets and the latter featuring 15 pairs of CVE code samples before and after patches, tested across four prompting strategies for a total of 120 cases. CTI-RCM includes 1,000 CVE descriptions from 2024, evaluating LLMs' threat intelligence analysis capabilities by assessing their accuracy in mapping vulnerabilities to their corresponding CWE classifications.} To highlight the effectiveness of the improvements made using CSEBenchmark, we choose three models from the relatively lower-performing Llama series---Llama-3.1-8B, Llama-3.1-70B, and Llama-3.2-3B---as subjects for enhancement. \newtext{Additionally, to assess whether high-performing LLMs can likewise benefit from these enhancements, we include GPT-4o in our experiments.}

First, we perform an initial evaluation of the original LLMs on the three assessment datasets and record instances where each model makes incorrect predictions. Next, we extract the knowledge gaps (i.e., knowledge points with an accuracy below 90\%) of each model from CSEBenchmark and provide this gap information to the LLMs for a reevaluation of the error instances. The proportion of previously incorrect predictions corrected in the reevaluation reflects the performance improvement of the LLMs after addressing their knowledge gaps. We employ a Retrieval-Augmented Generation (RAG) approach to inject the models with knowledge points related to their knowledge gaps. Specifically, for ease of implementation, we construct a vector database for each LLM using Milvus~\cite{2021milvus} and use corresponding question-answer pairs from CSEBenchmark to address the model's knowledge gaps. For embedding, we utilize the BGE-M3 model~\cite{bge-m3}. Before issuing the request to the LLMs, we \removedtext{match the top-5 most relevant entries from the vector database}\newtext{use each dataset's task instruction to query the vector database, retrieve the top-5 most relevant entries,} and incorporate them into the original prompt, \newtext{with the instruction, ``Please use the following retrieved context to answer the question,''} effectively addressing the models' knowledge gaps.

\begin{table}[htbp]
    \centering
    \footnotesize
    \caption{Performance improvement of LLMs after knowledge gap supplementation, \removedtext{with the numbers representing the percentage improvement in correcting previously incorrect predictions.}\newtext{with the numbers on either side of the arrow representing the count of error instances before and after enhancement. The percentages represent the proportion of previously incorrect instances that become correct after enhancement.}}
    \label{tab:improvement}
    \begin{tabular}{c|ccc}
    \hline
        \multirow{2}{*}{\textbf{Model}} & \multicolumn{3}{c}{\textbf{Benchmark}} \\ \cline{2-4}
            & \multicolumn{1}{c}{VuldetectBench} & SecLLMHolmes & CTI-RCM \\ \hline
        L3.2-3B & \removedtext{78.18\%$\uparrow$}\newtext{495$\rightarrow$108 (78\%)} & \removedtext{30.77\%$\uparrow$}\newtext{65$\rightarrow$45 (31\%)} & \removedtext{7.52\%$\uparrow$}\newtext{758$\rightarrow$701 (8\%)} \\ \hline
        L3.1-8B & \removedtext{84.18\%$\uparrow$}\newtext{373$\rightarrow$59 (84\%)} & \removedtext{25.42\%$\uparrow$}\newtext{59$\rightarrow$44 (25\%)} & \removedtext{14.75\%$\uparrow$}\newtext{434$\rightarrow$370 (15\%)} \\ \hline
        L3.1-70B & \removedtext{29.16\%$\uparrow$}\newtext{439$\rightarrow$311 (29\%)} & \removedtext{24.24\%$\uparrow$}\newtext{66$\rightarrow$50 (24\%)} & \removedtext{10.00\%$\uparrow$}\newtext{350$\rightarrow$315 (10\%)} \\ \hline
        \newtext{GPT-4o} & \newtext{405$\rightarrow$343 (15\%)} & \newtext{73$\rightarrow$55 (25\%)} & \newtext{248$\rightarrow$226 (9\%)} \\ \hline 
    \end{tabular}
\end{table}

The results in Table~\ref{tab:improvement} show that after addressing the knowledge gaps, all LLMs show improvements across the three datasets, confirming that the knowledge gaps identified by CSEBenchmark enhance LLM performance, with the highest improvement reaching \removedtext{84.18}\newtext{84}\%. For example, C++ is a knowledge gap for both Llama-3.2-3B and Llama-3.1-8B. The question-answer pairs on pointer operations within the knowledge points, such as \textit{``What should you do to a pointer after deleting the memory it points to, to avoid dangling pointer issues? Set the pointer to nullptr''} and \textit{``To ensure a reference cannot change the bound object, which declaration is appropriate? const int \&cri = i''}, help the models better understand the concept of pointer safety, which in turn enable them to correctly identify potential vulnerabilities related to improper pointer operations and memory deallocation in code. \newtext{Similarly, in CTI-RCM, RAG improves XSS vulnerability classification by providing definitions, enhancing model performance. Furthermore, we find that the retrieved semantically relevant question-answer pairs from the model's entire knowledge gap may not always precisely match the required knowledge but still contribute to overall performance improvement. For instance, a Go-related null pointer dereferencing question-answer pair helps the model identify a C++ null pointer dereferencing vulnerability in VulDetectBench.} Note that RAG technique used in this study is straightforward, and optimizing its design in the future could further enhance LLM performance.


\begin{findingbox}
    The knowledge gaps identified by CSEBenchmark can be used to improve model performance.
\end{findingbox}

\subsection{LLM Job Role Assessment (RQ4)}
\label{subsec:job_assessment}

Although we assess the selected LLMs on 345 knowledge points, real-world cybersecurity roles typically do not require proficiency at all of these points (though more coverage is generally beneficial). To evaluate how well these \removedtext{LLMs align}\newtext{LLMs' knowledge aligns} with the specific requirements of real-world cybersecurity positions, we gather job requirements from companies such as Amazon, Google, and Microsoft. Based on role descriptions, we manually map these requirements to our knowledge points. For example, the Amazon Security Engineer role specifies a requirement for \textit{``experience with a focus in areas such as systems, network, and/or application security.''} Drawing on our own expertise, we map this requirement to relevant CSEBenchmark knowledge points in system security (e.g., \textit{Linux security concepts}), network security (e.g. \textit{DoS vs DDoS}), and application security (e.g., \textit{Web Based Attacks and OWASP10}) to assess each LLM's alignment with the core skills needed for this role. In total, we identify six distinct roles for the analysis: Google's Senior Intelligence Analyst and Red Team Security Consultant, Amazon's Privacy Engineer, ISC Security Engineer, and Security Engineer, and Microsoft's Red Team Security Engineer. \newtext{The mapped knowledge points for each job role is provided in Appendix~\ref{appendix:job_map}.}

\begin{figure}[htbp]
    \centering
    \includegraphics[width=1.0\linewidth]{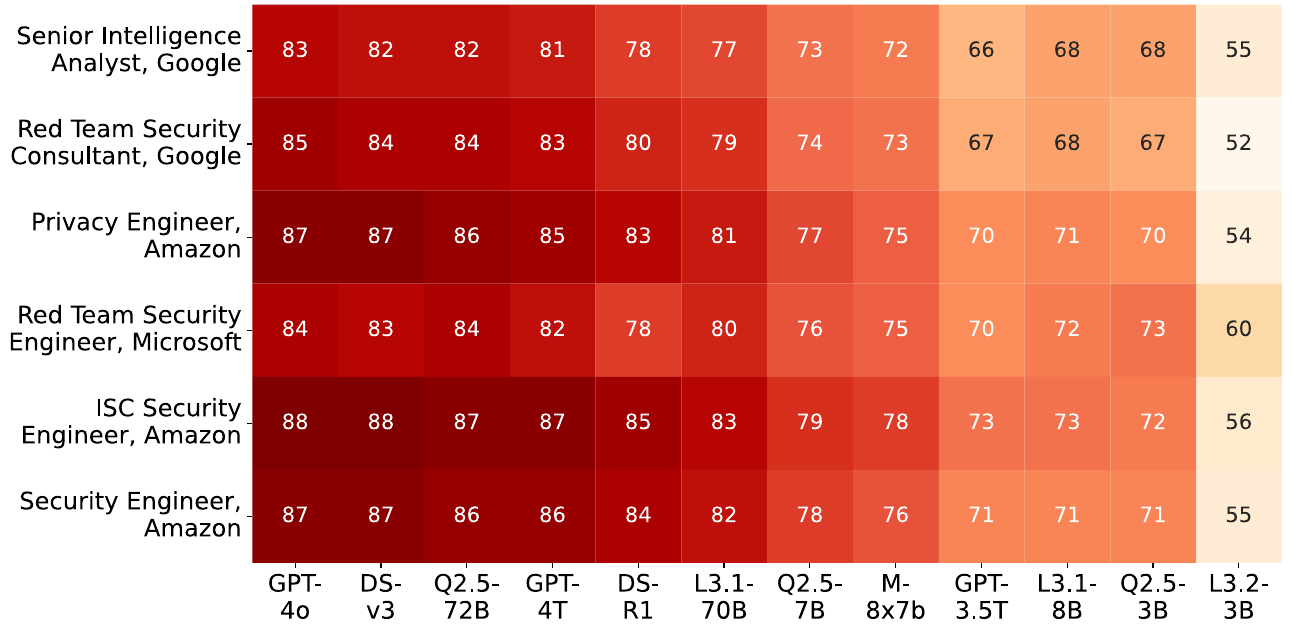}
    \caption{Heatmap of selected LLMs' match scores across six real-world cybersecurity job roles\removedtext{ (Columns from Left to Right: GPT-4o, Qwen-2.5-72B, GPT-4-Turbo, Llama-3.1-70B, Qwen-2.5-7B, Mixtral-8x7B, GPT-3.5-Turbo, Llama-3.1-8B, Qwen-2.5-3B, Llama-3.2-3B)}.}
    \label{fig:job_scores}
\end{figure}

\begin{figure*}[htbp]
    \centering
    \includegraphics[width=0.7\textwidth]{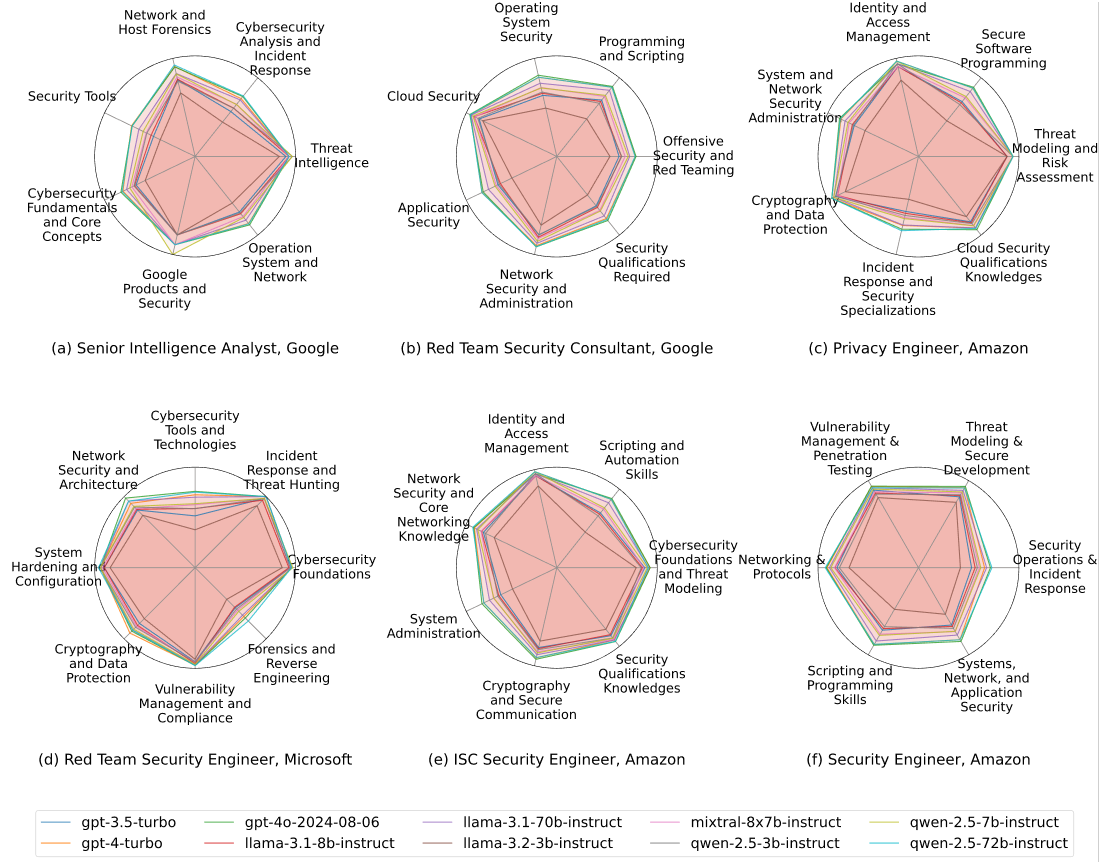}
    \caption{Radar chart showing the alignment of the selected LLMs with the requirements of six real-world cybersecurity job roles.}
    \label{fig:job_model}
\end{figure*}

We calculate the overall accuracy of mapped knowledge points as the job-role match score, with the results shown in Figure~\ref{fig:job_scores}. The ranking is closely aligned with the performance of each model on the CSEBenchmark. GPT-4o achieves the highest \newtext{knowledge} match scores for  \removedtext{the Google Red Team Security Consultant and the Amazon Privacy Engineer, ISC Security Engineer, and Security Engineer roles.}\newtext{the Google Senior Intelligence Analyst and the Google Red Team Security Consultant.} \newtext{For the Amazon Privacy Engineer, ISC Security Engineer, and Security Engineer, Deepseek-V3 and GPT-4o share the top position.} Similarly, for the Microsoft Red Team Security Engineer role, \removedtext{Qwen-2.5-72B achieves the best match, while for the Google Senior Intelligence Analyst,} Qwen-2.5-72B and GPT-4o are tied for first place. Notably, \newtext{knowledge} match scores for even the highest ranked LLMs are below 90\%, indicating that these models still do not fully meet the real-world cybersecurity job requirements.

\begin{findingbox}
    LLMs show limited \newtext{knowledge} alignment with cybersecurity job roles in the real world, with the highest match below 90\%. Different LLMs exhibit unique strengths aligned with specific roles.
\end{findingbox}

In addition, we group the required competencies for each role into core categories based on job descriptions and create radar charts to visually highlight current gaps for LLMs in each position, as shown in Figure~\ref{fig:job_model}. For the Google Senior Intelligence Analyst role, gaps appear in \textit{Cybersecurity Analysis and Incident Response} and \textit{Security Tools}, with top match scores of 77 and 70, respectively. For the Google Red Team Consultant role, the main gap is in \textit{Offensive Security and Red Teaming}, with a maximum score of 79. The roles of the Amazon Privacy Engineer and Security Engineer show gaps in \textit{Incident Response and Security Specializations} and \textit{Security Operations and Incident Response}, with top scores of 76 and 73, respectively. For the Amazon ISC Security Engineer role, LLMs perform more consistently, with scores above 80 across all areas. The Microsoft Red Team Security Engineer role highlights gaps in \textit{Cybersecurity Tools and Technologies} and \textit{Forensics and Reverse Engineering}, with highest scores of 76 and 75.

\begin{findingbox}
    Different cybersecurity roles reveal unique competency gaps for LLMs.
\end{findingbox}

\section{Discussion}
\label{sec:discussion}

\subsection{\newtext{Potential Cyclical Use and Model Bias}}
\label{subsec:bias}

\newtext{We observe that GPT-4-Turbo generates and answers its own questions, which can be seen as cyclical use. However, this does not impact the results in our paper. For other models, no cyclical use occurs, ensuring the validity of their results. Note that GPT-4-Turbo and GPT-4o are distinct models with different training data and methodologies. For GPT-4-Turbo, we believe there is no ``unfair cyclical use,'' as our carefully designed prompts ensure it solely relies on its summarization capabilities rather than its internal knowledge. To verify this, human experts manually examine 500 randomly selected questions to identify their corresponding source passages within the corpus. The process involves first identifying potential passages by searching for distinctive keywords in each question, followed by a thorough analysis to determine whether the passages contained all key concepts relevant to the question. A passage is considered the source if it fully encompasses these key concepts. In all cases, a corresponding passage is found, confirming that GPT-4-Turbo generates questions exclusively based on the provided material. Since the corpus is not available when answering the questions, no unfair cyclical use occurrs, ensuring the credibility of the results.}

\newtext{Additionally, considering the possibility that GPT-4-Turbo might introduce its own preferences when generating questions, potentially leading to bias, we conduct an evaluation to assess topic selection fairness. We randomly select three distinct knowledge points (\textit{Kerberos}, \textit{Packet Sniffer}, and \textit{Nikto}) and asked GPT-4-Turbo, GPT-4o, Llama-3.1-70B, and Qwen-2.5-72B to extract topics from the corpus. These topics directly influence the question distribution, as five questions will be generated for each topic. Therefore, any skew in the topic distribution can reflect potential model bias. The topic distribution in semantic space (via BGE-M3) shows consistent results across the four models, with no bias observed (see Appendix~\ref{appendix:topic_distribution}).}

\subsection{Limitation and Future Work}
\label{subsec:limitaion}

Despite undergoing 772 hours of manual review and correction, the CSEBenchmark still presents certain limitations. First, our knowledge framework, based on three public cybersecurity roadmaps, covers 345 knowledge points of cybersecurity experts. However, some specialized areas, such as hardware security, may be underrepresented. To improve the framework, we plan to expand the knowledge points through interviews with cybersecurity professionals, ensuring that it addresses emerging needs. Second, each knowledge point question in the CSEBenchmark is generated based on a single, most relevant, and official source (e.g., textbooks, reputable websites, or blogs), providing a degree of reliability. However, a single source may sometimes fail to comprehensively cover the full scope of a knowledge point. We plan to address this by supplementing each knowledge point with additional relevant materials. Third, in our evaluation, we employ three commonly used prompting methods---Zero-shot, Few-shot, and CoT---to probe the upper knowledge limits of LLMs, using the highest score as the final result to reveal critical knowledge gaps. However, in practical applications, more advanced prompting techniques may further improve LLM performance, and we aim to incorporate such advanced techniques for a more thorough assessment of LLM capabilities. Lastly, CSEBenchmark relies on \textit{xFinder} as the back-end technology to extract answers from free text. Compared to regex-based methods, xFinder provides substantial accuracy improvements; however, sampling indicates that an error rate of 8\% persists. To ensure fair and objective evaluation outcomes, it is necessary to further enhance xFinder's accuracy in future work.

\noindent\textbf{Impact of Time on Evaluation Results.} Due to varying knowledge cutoff dates, some newer source materials may only appear in the training data of models with later cutoffs. However, our objective is to highlight existing knowledge gaps in LLMs---gaps that may stem from limited training or incomplete data. These gaps are objectively present, regardless of the cause, making discussions on knowledge cut-off dates secondary. Our focus remains on objectively identifying and analyzing these gaps to accurately assess the practical limits of LLM capabilities in cybersecurity. Furthermore, with the rapid evolution of LLMs, the conclusions in this study may become outdated over time. Continued evaluation is essential to answer the question, \textit{``how far have we come in achieving a digital cybersecurity expert?''} and to ensure that our findings reflect the latest advances and changes in LLM capabilities.
\section{Conclusion}
\label{sec:conclusion}

To assess the knowledge gaps in LLMs in fulfilling the role of a digital security expert, this study develops a cybersecurity knowledge model based on cognitive science, encompassing 345 fine-grained knowledge points, and constructs a benchmark dataset, CSEBenchmark, containing 11,050 questions. Evaluation across \removedtext{10}\newtext{12} popular LLMs reveals that their overall accuracy is currently limited to 85.42\%, with notable gaps in specialized procedural knowledge, such as the use of professional tools and uncommon commands. Additionally, different LLMs have unique knowledge gaps, and even larger models within the same family may underperform on certain knowledge points where smaller models perform better. By addressing these knowledge gaps, we achieve up to an \removedtext{84.18}\newtext{84}\% improvement in correcting previously incorrect predictions across three benchmarks for two cybersecurity tasks, thereby validating the effectiveness of our findings.
\ifCLASSOPTIONcompsoc
  \section*{Acknowledgments}
\else
  \section*{Acknowledgment}
\fi

We are grateful to our shepherd and the anonymous reviewers for their valuable guidance and insightful comments. This research is supported by Zhongguancun Laboratory and the Beijing Outstanding Young Scientist Program (No. JWZQ20240101008).

\bibliographystyle{IEEEtran}
\bibliography{references}
\appendices

\section{Prompts for Answer Generation}
\label{appendix:prompt_answer_generation}

\begin{figure}[htbp]
    \centering
    \includegraphics[width=1.0\linewidth]{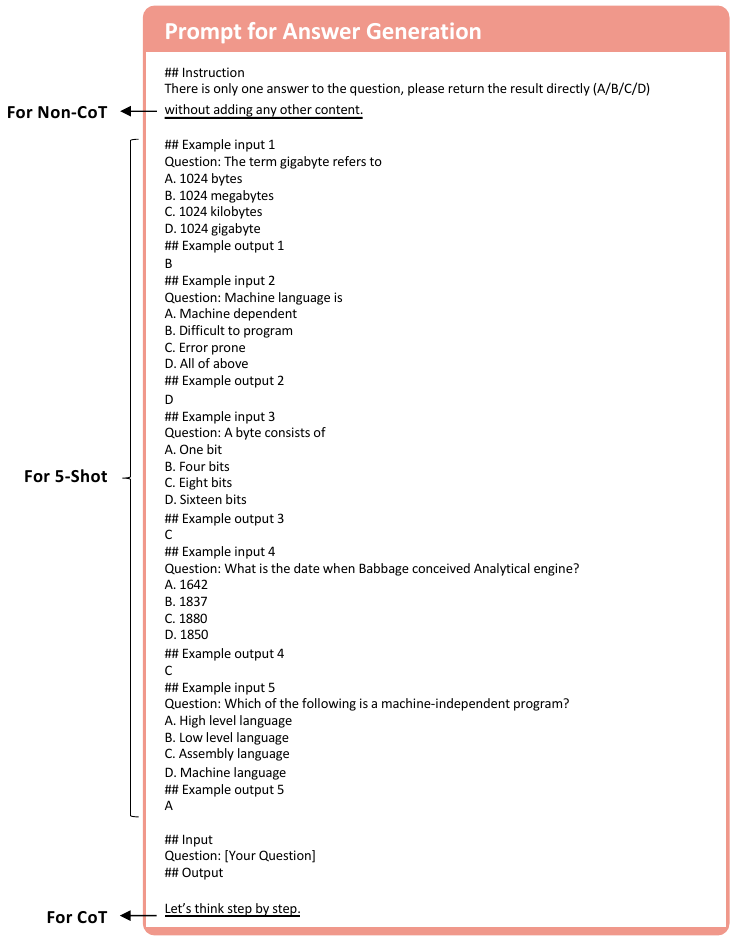}
    \caption{Prompt for answer genenration.}
    \label{fig:prompt_answer_generation}
\end{figure}

Figure~\ref{fig:prompt_answer_generation} presents the prompt used for generating answers.

\section{\newtext{Knowledge Points for Each Job Role}}
\label{appendix:job_map}

\newtext{Due to space limitations, the mapped knowledge points for each job role are provided in our repository: \url{https://github.com/NASP-THU/CSEBenchmark}}

\section{\newtext{Topic Distribution in Semantic Space}}
\label{appendix:topic_distribution}

\begin{figure}[htbp]
    \centering
    \includegraphics[width=1.0\linewidth]{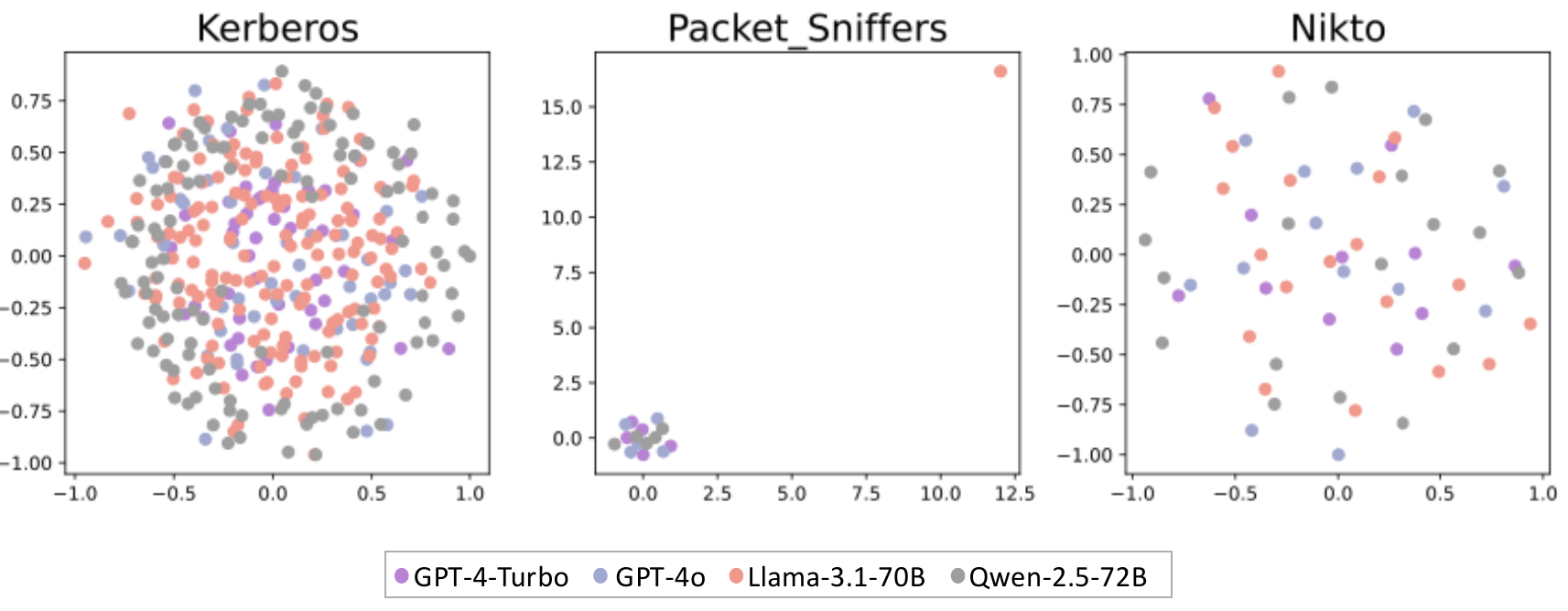}
    \caption{\newtext{Topic distribution in semantic space across GPT-4-Turbo, GPT-4o, Llama-3.1-70B, and Qwen-2.5-72B.}}
    \label{fig:topic_distribution}
\end{figure}

\newtext{Figure~\ref{fig:topic_distribution} presents the topic distribution in semantic space across LLMs.}

\clearpage

\section{Meta-Review}
\label{appendix:meta-review}

The following meta-review was prepared by the program committee for the 2025 IEEE Symposium on Security and Privacy (S\&P) as part of the review process as detailed in the call for papers.

\subsection{Summary}

This paper proposes a new benchmark, CSEBenchmark, to evaluate the cybersecurity knowledge of Large Language Models. CSEBenchmark contains 11,050 questions, covering three types of knowledge: factual knowledge (to be memorized), conceptual knowledge (requiring understanding of underlying principles), and procedural knowledge (requiring hands-on practice). To construct the benchmark, it took 672 man-hours of reviewing the LLM-generated questions and 100 man-hours of corrections.

\subsection{Scientific Contributions}

\begin{itemize}
    \item Independent Confirmation of Important Results with Limited Prior Research
    \item Provides a New Data Set For Public Use
    \item Provides a Valuable Step Forward in an Established Field
\end{itemize}

\subsection{Reasons for Acceptance}

\begin{enumerate}
    \item This paper has independently confirmed important results with limited prior research. The paper demonstrates that having cybersecurity knowledge can significantly boost the performance of vulnerability detection and threat intelligence analysis tasks, via retrieval-augmented generation (RAG)
    \item This paper provides a new data set for public use. CSEBenchmark enables a fine-grained and detailed evaluation of LLMs on cybersecurity knowledge.
    \item This paper provides a valuable step forward in an established field. The paper provides comprehensive evaluation of cybersecurity expertise in popular LLM models and identifying their knowledge gaps in this area.
\end{enumerate}

\subsection{Noteworthy Concerns}

\begin{enumerate}
    \item The dataset could be biased since only GPT-4-Turbo is used to generate the dataset. It might be more reasonable to use different LLMs to generate questions, combined with manual verification.
    \item It is unclear whether the proposed evaluation framework approximates expert level knowledge of human security analysts.
\end{enumerate}

\section{Response to the Meta-Review}
\label{appendix:meta-review_response}

\noindent\textbf{Response to concern 1.} Thank you for pointing out this issue. We discuss the impact of cyclical use in Section~\ref{subsec:bias}. In future work, we will explore using other advanced LLMs such as DeepSeek and Claude in GPT-4-Turbo's current role in question generation, enabling a more comprehensive evaluation through cross-model question generation and answering.

\noindent\textbf{Response to concern 2.} We acknowledge the importance of validating whether the proposed evaluation framework approximates expert-level knowledge of human security analysts. However, given the time and cost constraints of traditional expert surveys, we are currently unable to conduct such an experiment. Nonetheless, we believe our study provides a valuable benchmark for assessing LLM performance in cybersecurity tasks, and future work could incorporate expert evaluations to further refine the framework.

\end{document}